\documentclass[conference]{IEEEtran}
\usepackage{fancyhdr}
\usepackage{ifpdf}
\usepackage{verbatim}
\usepackage{cite}
\ifCLASSINFOpdf
   \usepackage[pdftex]{graphicx}
  \DeclareGraphicsExtensions{.pdf,.jpeg,.png}
\else
  \usepackage[dvips]{graphicx}
\fi
\usepackage{algorithmic}
\usepackage{algorithm}
\usepackage{listings}
\usepackage{array}
\usepackage{mdwmath}
\usepackage{mdwtab}
\usepackage{eqparbox}
\usepackage[caption=false,font=footnotesize]{subfig}
\usepackage{fixltx2e}
\usepackage{stfloats}
\usepackage{url}
\begin{document}

\title{\large{\bf The Universe at Extreme Scale: Multi-Petaflop 
Sky Simulation on the BG/Q}}

\author{\IEEEauthorblockN{Salman Habib, Vitali Morozov, Hal Finkel, Adrian
  Pope, Katrin Heitmann, Kalyan Kumaran, Tom Peterka,  Joe Insley\\}
\IEEEauthorblockA{Argonne National Laboratory \\
\{habib, morozov, hfinkel, apope, heitmann, kumaran, tpeterka, insley\}@anl.gov}
\and
\IEEEauthorblockN{David Daniel, Patricia Fasel, Nicholas Frontiere \\ }
\IEEEauthorblockA{Los Alamos National Laboratory \\
\{ddd, pkf, frontiere\}@lanl.gov}
\and
\IEEEauthorblockN{Zarija Luki\'c \\ }
\IEEEauthorblockA{Lawrence Berkeley National Laboratory \\
\{zarija\}@lbl.gov}
}

\maketitle
\thispagestyle{fancy}
\lhead{}
\rhead{}
\chead{}
\rfoot{}
\cfoot{}
\renewcommand{\headrulewidth}{0pt}
\renewcommand{\footrulewidth}{0pt}

\begin{abstract}
  Remarkable observational advances have established a compelling
  cross-validated model of the Universe. Yet, two key pillars of this
  model -- dark matter and dark energy -- remain mysterious. Sky
  surveys that map billions of galaxies to explore the `Dark Universe',
  demand a corresponding extreme-scale simulation capability; the HACC
  (Hybrid/Hardware Accelerated Cosmology Code) framework has been
  designed to deliver this level of performance now, and into the
  future. With its novel algorithmic structure, HACC allows flexible
  tuning across diverse architectures, including accelerated and
  multi-core systems.

  On the IBM BG/Q, HACC attains unprecedented scalable performance --
  currently 13.94 PFlops at 69.2\% of peak and 90\% parallel
  efficiency on 1,572,864 cores with an equal number of MPI ranks, and
  a concurrency of 6.3 million. This level of performance was achieved
  at extreme problem sizes, including a benchmark run with more than
  3.6 trillion particles, significantly larger than any cosmological
  simulation yet performed.
\end{abstract}

\section{introduction}
\label{sec:intro}

Modern cosmology is one of the most exciting areas in physical
science. Decades of surveying the sky have culminated in a
cross-validated, ``Cosmological Standard Model''. Yet, key pillars of
the model -- dark matter and dark energy -- together accounting for
95\% of the Universe's mass-energy remain
mysterious~\cite{frieman,DEreview}. Deep fundamental questions demand
answers: What is the dark matter? Why is the Universe's expansion
accelerating?  What is the nature of primordial fluctuations?  Should
general relativity be modified?
\begin{figure}[t!]
  \centering \includegraphics[width=2.7in,angle=0]{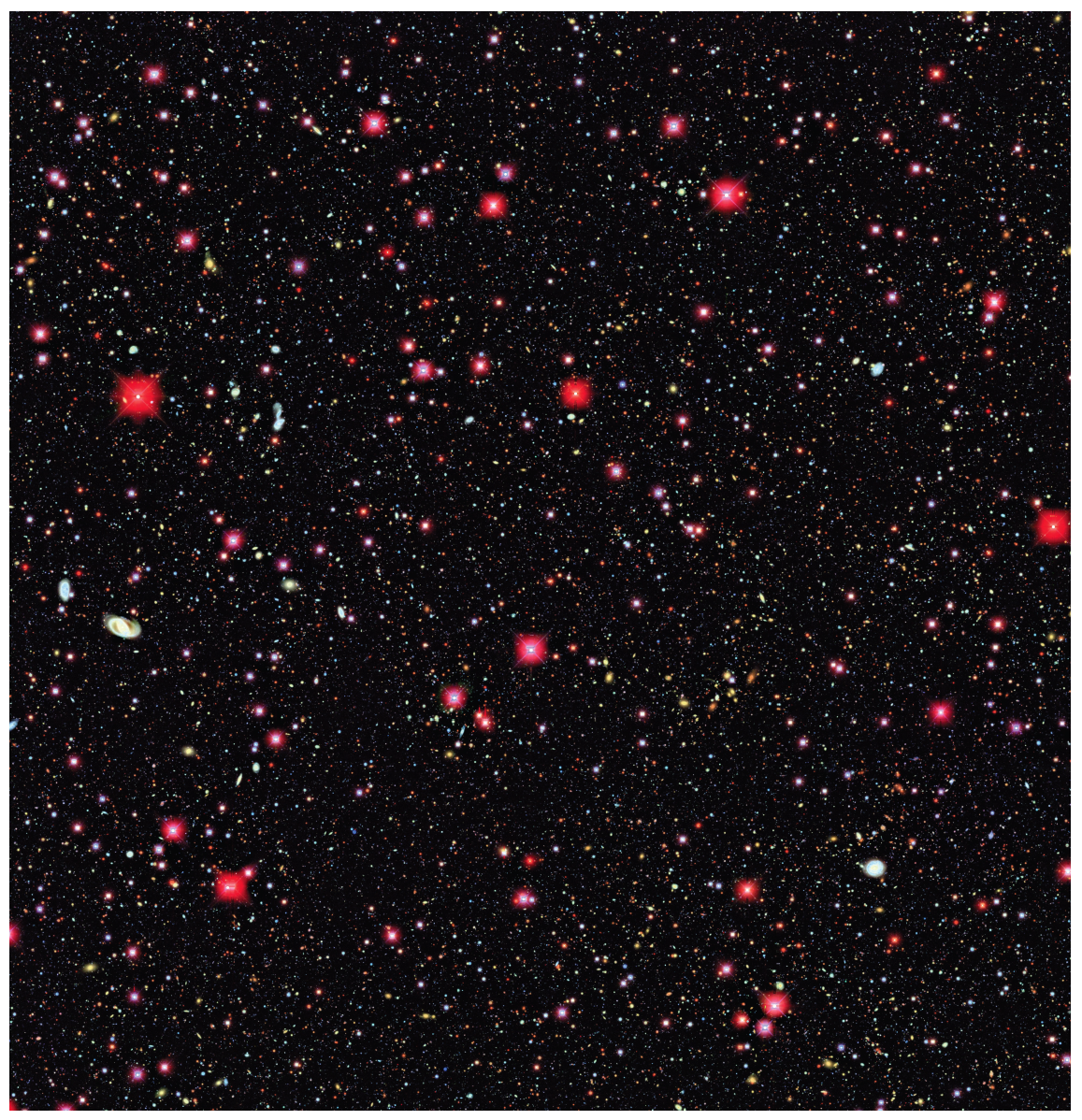}

\vspace{-0.2cm}

\caption{\small{\em Deep Lens Survey~\cite{wittman} image of galaxies
    covering a patch of the sky roughly equivalent to the size of the
    full moon. LSST will not only go deeper but it will cover 50,000
    times the area of this image in 6 optical wavelength bands.}}
\label{dls}
\end{figure}

To address these questions, ground and space-based observatories
operating at multiple wavebands~\cite{stevemyers} are aiming to unveil
the true nature of the ``Dark Universe''. Driven by advances in
semiconductor technology, surveys follow a version of Moore's law, in
terms of CCD pixels or surveyed galaxies per year. In a major leap
forward, current cosmological constraints will soon be improved by an
order of magnitude~\cite{detf}. As an example, the Large Synoptic
Survey Telescope (LSST)~\cite{lsst} can be compared to today's
observations from the Sloan Digital Sky Survey (SDSS)~\cite{sdss}: In
one night LSST will capture data equivalent to five years of SDSS
imaging (Fig.~\ref{dls})!

Interpreting future observations will be impossible without a modeling
and simulation effort as revolutionary as the new surveys: The desired
size and performance improvements for simulations over the next decade
are measured in orders of magnitude~\cite{doe_hep_rep}. Because the
simulations to be run are memory-limited on even the largest machines
available and a large number of them are necessary, very stringent
requirements are simultaneously imposed on code performance and
efficiency. We show below how HACC meets these exacting conditions by
attaining unprecedented sustained levels of performance, reaching up
to $69\%$ of peak on certain BG/Q partition sizes.

Cosmic structure formation is described by the gravitational
Vlasov-Poisson equation in an expanding Universe~\cite{peebles}, a 6-D
PDE for the Liouville flow (\ref{le}) of the phase space PDF where
self-consistency is imposed by the Poisson equation (\ref{pe}):
\begin{eqnarray}
&&\partial_t f({\mathbf{x}},
{\mathbf{p}})+\dot{\mathbf{x}}\cdot\partial_{\mathbf{x}}
f({\mathbf{x}}, {\mathbf{p}}) -
\nabla\phi\cdot\partial_{\mathbf{p}}f({\mathbf{x}}, 
{\mathbf{p}})=0,\label{le}~~~~~~\\
&&\nabla^2\phi({\mathbf{x}})= 4\pi
Ga^2(t)\Omega_m\delta_m({\mathbf{x}})\rho_c.
\label{pe}
\end{eqnarray}
The expansion of the Universe is encoded in the time-dependence of the
scale factor $a(t)$ governed by the cosmological model, the Hubble
parameter, $H=\dot{a}/a$, $G$ is Newton's constant, $\rho_c$ is the
critical density, $\Omega_m$, the average mass density as a fraction
of $\rho_c$, $\rho_m({\mathbf{x}})$ is the local mass density, and
$\delta_m({\mathbf{x}})$ is the dimensionless density contrast,
\begin{eqnarray}
&&\rho_c=3H^2/8\pi G,~~~
\delta_m({\mathbf{x}})=(\rho_m({\mathbf{x}})-
\langle\rho_m\rangle)/\langle\rho_m\rangle,~~~~\\   
&&{\mathbf{p}}=a^2(t) \dot{\mathbf{x}},
~~~\rho_m({\mathbf{x}})= a(t)^{-3}m\int
d^3{\mathbf{p}}f({\mathbf{x}}, {\mathbf{p}}). 
\label{defs}
\end{eqnarray}
The Vlasov-Poisson equation is very difficult to solve directly
because of its high dimensionality and the development of structure --
including complex multistreaming -- on ever finer scales, driven by the
gravitational Jeans instability. Consequently, N-body methods, using
tracer particles to sample $f({\mathbf{x}}, {\mathbf{p}})$ are used;
the particles follow Newton's equations in an expanding Universe, with
the forces given by the gradient of the scalar potential as computed
from Eq.~(\ref{pe})~\cite{nbodyrev}.

Under the Jeans instability, initial perturbations given by a smooth
Gaussian random field evolve into a `cosmic web' comprising of
sheets, filaments, and local mass concentrations called
halos~\cite{cosweb, mwhite}. The first stars and galaxies form in
halos and then evolve as the halo distribution also evolves by a
combination of dynamics, mass accretion and loss, and by halo
mergers. To capture this complex behavior, cosmological N-body
simulations have been developed and refined over the last three
decades~\cite{nbodyrev}. In addition to gravity, gasdynamic, thermal,
radiative, and other processes must also modeled, e.g., sub-grid
modeling of star formation. Large-volume simulations usually
incorporate the latter effects via semi-analytic modeling.

\begin{figure}{}
  \centering \includegraphics[width=2.9in,angle=0]{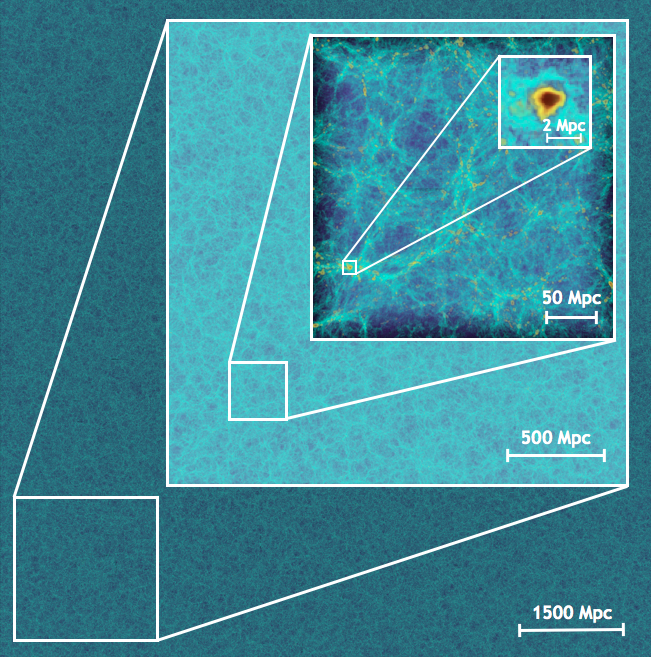}

\vspace{-0.2cm}

\caption{\small{\em Zoom-in visualization of the density field in a
    1.07 trillion particle, $9.14~\hbox{Gpc}$ box-size simulation with
    HACC on sixteen BG/Q racks. This figure illustrates the global
    spatial dynamic range covered by the simulation, $\sim 10^6$,
    although the finer details are not resolved by the
    visualization. Simulation details are covered in
    Section~\ref{science}.}}
\label{zoom}
\end{figure}

To understand the essential nature of the challenge posed by future
surveys, a few elementary arguments suffice. Survey depths are of
order a few Gpc (1~${\rm pc}=3.26$~light-years); to follow typical
galaxies, halos with a minimum mass of $\sim$$10^{11}$~M$_\odot$
($\hbox{M}_\odot=1$~solar mass) must be tracked. To properly resolve
these halos, the tracer particle mass should be
$\sim$$10^{8}$~M$_\odot$ and the force resolution should be small
compared to the halo size, i.e., $\sim$kpc. This last argument
immediately implies a dynamic range (ratio of smallest resolved scale
to box size) of a part in $10^6$ ($\sim$Gpc/kpc) everywhere in the
{\em entire} simulation volume (Fig.~\ref{zoom}). The mass resolution
can be specified as the ratio of the mass of the smallest resolved
halo to that of the most massive, which is $\sim$$10^5$. In terms of
the number of simulation particles, this yields counts in the range of
hundreds of billions to trillions. Time-stepping criteria follow from
a joint consideration of the force and mass
resolution~\cite{power}. Finally, stringent requirements on accuracy
are imposed by the very small statistical errors in the observations
-- certain quantities such as lensing shear power spectra must be
computed at accuracies of a {\em fraction} of a
percent~\cite{lensing}.
\noindent

For a cosmological simulation to be considered ``high-resolution'',
{\em all} of the above demands must be met. In addition, throughput is
a significant concern. Scientific inference from sets of cosmological
observations is a statistical inverse problem where many runs of the
forward problem are needed to obtain estimates of cosmological
parameters via Markov chain Monte Carlo methods. For many analyses,
hundreds of large-scale, state of the art simulations will be
required~\cite{ccf}.

The structure of the HACC framework is based on the realization that
it must not only meet the challenges of spatial dynamic range, mass
resolution, accuracy, and throughput, but also overcome a final
hurdle, i.e., be fully cognizant of coming disruptive changes in
computational architectures. As a validation of its design philosophy,
HACC was among the pioneering applications proven on the heterogeneous
architecture of Roadrunner~\cite{hacc1,hacc2}, the first supercomputer
to break the petaflop barrier.

HACC's multi-algorithmic structure also attacks several weaknesses of
conventional particle codes including limited vectorization,
indirection, complex data structures, lack of threading, and short
interaction lists. It combines MPI with a variety of local programming
models (OpenCL, OpenMP) to readily adapt to different
platforms. Currently, HACC is implemented on conventional and
Cell/GPU-accelerated clusters, on the Blue Gene architecture, and is
running on prototype Intel MIC hardware. HACC is the first, and
currently the only large-scale cosmology code suite world-wide, that
can run at scale (and beyond) on {\em all} available supercomputer
architectures.

To showcase this flexibility, we present scaling results for two
systems aside from the BG/Q in Section~\ref{sec:results}; on the
entire ANL BG/P system and over all of Roadrunner. Recent HACC science
results on Roadrunner include a suite of 64~billion particle runs for
baryon acoustic oscillations predictions for BOSS (Baryon Oscillation
Spectroscopic Survey)~\cite{boss_bao} and a high-statistics study of
galaxy cluster halo profiles~\cite{conc_hacc}.

HACC's performance and flexibility are not dependent on
vendor-supplied or other high-performance libraries or linear algebra
packages; the 3-D parallel FFT implementation in HACC couples high
performance with a small memory footprint as compared to available
libraries. Unlike some other high-performance N-body codes, HACC does
not use any special hardware. The implementation for the BG/Q
architecture has far more generally applicable features than (the HACC
or other) CPU/GPU short-range force implementations.

\section{HACC Framework: General Features}
\label{sec:hacc}

The cosmological N-body problem is typically treated by a mix of grid
and particle-based techniques. The HACC design accepts that, as a
general rule, particle and grid methods both have their
limitations. For physics and algorithmic reasons, grid-based
techniques are better suited to larger (`smooth') lengthscales, with
particle methods having the opposite property. This suggests that
higher levels of code organization should be grid-based, interacting
with particle information at a lower level of the computational
hierarchy.

\begin{figure}{}
  \centering \includegraphics[width=2.4in,angle=0]{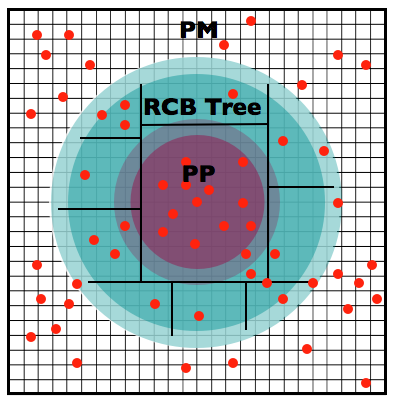}

\vspace{-0.2cm}

\caption{\small{\em Informal representation of the HACC force
    evaluation hierarchy -- 1) long/medium-range contributions from a
    high-order grid-based, spectrally filtered particle-mesh (PM)
    solver, 2) medium/short-range contributions using a (rank-local)
    recursive coordinate bisection (RCB) tree algorithm (green
    region), 3) close-range contributions using direct
    particle-particle (PP) interactions (magenta). Parameters
    governing the cross-overs are discussed in the text.}}
\label{haccforce}
\end{figure}

Following this central idea, HACC uses a hybrid parallel algorithmic
structure, splitting the gravitational force calculation into a
specially designed grid-based long/medium range spectral particle-mesh
(PM) component that is common to all architectures, and an
architecture-tunable particle-based short/close-range solver
(Fig.~\ref{haccforce}). The grid is responsible for 4 orders of
magnitude of dynamic range, while the particle methods handle the
critical 2 orders of magnitude at the shortest scales where particle
clustering is maximal and the bulk of the time-stepping computation
takes place.

The computational complexity of the PM algorithm~\cite{hockney} is
${\mathcal{O}}(N_p)$+${\mathcal{O}}(N_g\log N_g)$, where $N_p$ is the
total number of particles, and $N_g$ the total number of grid
points. The short-range tree algorithms~\cite{tree} in HACC can be
implemented in ways that are either ${\mathcal{O}}(N_{pl}\log N_{pl})$
or ${\mathcal{O}}(N_{pl})$, where $N_{pl}$ is the number of particles
in individual spatial domains ($N_{pl}\ll N_p$), while the close-range
force computations are ${\mathcal{O}}(N_d^2)$ where $N_d$ is the
number of particles in a tree leaf node within which all direct
interactions are summed. $N_d$ values can range from $\sim200$ in a
`fat leaf' tree, to as large as $10^{5}$ in the case of a CPU/GPU
implementation (no mediating tree).

HACC uses mixed precision computation -- double precision is used for
the spectral component of the code, whereas single precision is
adequate for the short/close-range particle force evaluations and
particle time-stepping. 

\noindent

HACC's long/medium range algorithm is based on a fast, spectrally
filtered PM method. The density field is generated from the particles
using a Cloud-In-Cell (CIC) scheme~\cite{hockney}, but is then
smoothed with the (isotropizing) spectral filter
\begin{equation}
\exp{(-k^2\sigma^2/4)}\left[(2k/\Delta)\sin(k\Delta/2)\right]^{n_s},
\label{filter}
\end{equation}
with the nominal choices $\sigma=0.8$, $n_s=3$. This reduces
the anisotropy ``noise'' of the CIC scheme by over an order of
magnitude without requiring complex and inflexible higher-order
spatial particle deposition methods. The noise reduction allows
matching the short and longer-range forces at a spacing of 3 grid
cells, with important ramifications for performance.

The Poisson solver uses a sixth-order, periodic, influence function
(spectral representation of the inverse Laplacian)~\cite{sh6}. The
gradient of the scalar potential is obtained using higher-order
spectral differencing (fourth-order Super-Lanczos~\cite{hamming}). The
``Poisson-solve'' in HACC is the composition of all the kernels above
in one single Fourier transform; each component of the potential field
gradient then requires an independent FFT. HACC uses its own scalable,
high performance 3-D FFT routine implemented using a 2-D pencil
decomposition (details are given in Section~\ref{sec:results}.)

To obtain the short-range force, the filtered grid force is subtracted
from the exact Newtonian force. The filtered grid force was obtained
numerically to high accuracy using randomly sampled particle pairs and
then fitted to an expression with the correct large and small distance
asymptotics. Because this functional form is needed only over a small,
compact region, it can be simplified using a fifth-order polynomial
expansion to speed up computations in the main force kernel
(Section~\ref{sec:bg}).

\begin{figure}[b!]
  \centering \includegraphics[width=2.3in,angle=0]{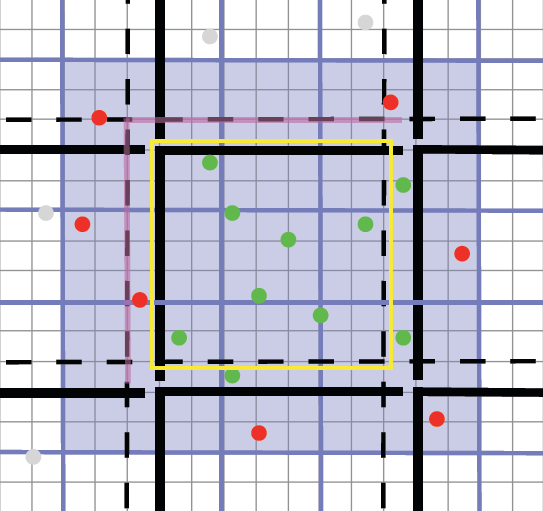}

\vspace{-0.2cm}

\caption{\small{\em Simplified 2-D sketch of HACC's 3-D
    particle overloading scheme. Thick black lines denote domain
    boundaries. Green particles lie within the central domain and are
    `active' -- their mass is deposited in the Poisson solve. The red
    particles are passive in the boundary regions of the central
    domain -- they are only moved by the force interpolated from the
    Poisson solver -- but (self-consistently) active in neighboring
    domains. Particles switch roles as they cross domain boundaries.}}
\label{overload}
\end{figure}

HACC's spatial domain decomposition is in regular (non-cubic) 3-D
blocks, but unlike the guard zones of a typical PM method, full
particle replication -- termed `particle overloading' -- is employed
across domain boundaries (Fig.~\ref{overload}). The typical memory
overhead cost for a large run is $\sim 10\%$. The point of overloading
is to allow essentially exact medium/long-range force calculations
with no communication of particle information and high-accuracy local
force calculations with relatively sparse refreshes of the overloading
zone (for details, see Ref.~\cite{hacc1}). The second advantage of
overloading is that it frees the local force solver from handling
communication tasks, which are taken care of by the long/medium-range
force framework. Thus new `on-node' local methods can be plugged in
with guaranteed scalability and only local optimizations are
necessary. Note that all short-range methods in HACC are local to the
MPI-rank and the locality can be fine-grained further. This locality
has the key advantage of lowering the number of levels in tree
algorithms and being able to parallelize across fine-grained particle
interaction sub-volumes.

The time-stepping in HACC is based on a 2nd-order split-operator
symplectic scheme that sub-cycles the short/close-range evolution
within long/medium-range `kick' maps where particle positions do not
change but the velocities are updated. The relatively slowly evolving
longer range force is effectively frozen during the shorter-range time
steps, which are a symmetric `SKS' composition of stream (position
update, velocity fixed) and kick maps for the short/close-range
forces~\cite{beams}:
\begin{equation}
M_{full}(t)=M_{lr}(t/2)(M_{sr}(t/n_c))^{n_c}M_{lr}(t/2).
\label{tstep}
\end{equation}
The number of sub-cycles can vary, depending on the force and mass
resolution of the simulation, from  $n_c=5-10$.

The long/medium-range solver remains unchanged across all
architectures. The short/close-range solvers are chosen and optimized
depending on the target architecture. These solvers can use direct
particle-particle interactions, i.e., a P$^3$M
algorithm~\cite{hockney}, as on Roadrunner, or use both tree and
particle-particle methods as on the IBM BG/P and BG/Q
(`PPTreePM'). The availability of multiple algorithms within the HACC
framework allows us to carry out careful error analyses, for example,
the P$^3$M and the PPTreePM versions agree to within $0.1\%$ for the
nonlinear power spectrum test in the code comparison suite of
Ref.~\cite{heitmann05}.

For heterogeneous systems such as Roadrunner, or in the near future,
Titan at OLCF, the long/medium-range spectral solver operates at the
CPU layer. Depending on the memory balance between the CPU and the
accelerator, we can choose to specify two different modes, 1) grids
held on the CPU and particles on the accelerator, or 2) a streaming
paradigm with grid and particle information primarily resident in CPU
memory with computations streamed through the accelerator. In both
cases, the local force solve is a direct particle-particle
interaction, i.e., the whole is a P$^3$M code with hardware
acceleration. For a many-core system, the top layer of the code
remains the same, but the short-range solver changes to a tree-based
algorithm which is much better suited to the Blue Gene and MIC
architectures. We will provide an in-depth description of our Blue
Gene/Q-specific implementation in Section~\ref{sec:bg}.

To summarize, the HACC framework integrates multiple algorithms and
optimizes them across architectures; it has several interesting
performance-enhancing features, e.g., overloading, spectral filtering
and differentiation, mixed precision, and compact local trees. HACC
attacks the weaknesses of conventional particle codes in ways made
fully explicit in the next Section -- lack of vectorization,
indirection, complex data structures, lack of threading, and short
interaction lists. Finally, weak scaling is a function only of the
spectral solver; HACC's 2-D domain decomposed FFT guarantees excellent
performance and scaling properties (see Section~\ref{sec:results}).

\section{HACC: BG/Q Implementation}
\label{sec:bg}

The BG/Q is the third generation of the IBM Blue Gene line of
supercomputers targeted primarily at large-scale scientific
applications, continuing the tradition of optimizing for price
performance, scalability, power efficiency, and system
reliability~\cite{redbook}. The new BG/Q Compute chip (BQC) is a
System-on-Chip (SoC) design combining CPUs, caches, network, and
messaging unit on a single chip~\cite{haring}. A single BG/Q rack
contains 1024 BG/Q nodes like its predecessors. Each node contains the
BQC and 16~GB of DDR3 memory.  Each BQC uses 17 augmented 64-bit
PowerPC A2 cores with specific enhancements for the BG/Q: 1) 4
hardware threads and a SIMD quad floating point unit (Quad Processor
eXtension, QPX), 2) a sophisticated L1 prefetching unit (L1P) with
both stream and list prefetching, 3) a wake-up unit to reduce certain
thread-to-thread interactions, and 4) transactional memory and
speculative execution. Of the 17 BQC cores, 16 are for user
applications and one for handling OS interrupts and other system
services. Each core has access to a private 16~KB L1 data cache and a
shared 32~MB multi-versioned L2 cache connected by a crossbar.  The A2
core runs at 1.6~GHz and the QPX allows for 4 FMAs per cycle,
translating to a peak performance per core of 12.8~GFlops, or
204.8~GFlops for the BQC chip.  The BG/Q network has a 5-D torus
topology; each compute node has 10 communication links with a peak
total bandwidth of 40~GB/s~\cite{chen}. The internal BQC interconnect
has a bisection bandwidth of 563~GB/s.

In order to evaluate the short-range force on non-accelerated systems,
such as the BG/Q, HACC uses a recursive coordinate bisection (RCB)
tree in conjunction with a highly-tuned short-range polynomial force
kernel. The implementation of the RCB tree, although not the force
evaluation scheme, generally follows the discussion in
Ref.~\cite{gafton}. Two core principles underlie the high performance
of the RCB tree's design.

{\em Spatial Locality.} The RCB tree is built by recursively dividing
particles into two groups. The dividing line is placed at the center
of mass coordinate perpendicular to the longest side of the box. Once
this line is chosen, the particles are partitioned such that particles
in each group occupy disjoint memory buffers. Local forces are then
computed one leaf node at a time. The net result is that the particle
data exhibits a high degree of spatial locality after the tree build;
because the computation of the short-range force on the particles in
any given leaf node, by construction, deals with particles only in
nearby leaf nodes, the cache miss rate during the force computation is
extremely low.

{\em Walk Minimization.} In a traditional tree code, an interaction
list is built and evaluated for each particle. While the interaction
list size scales only logarithmically with the total number of
particles (hence the overall ${\mathcal{O}}(N\log N)$ complexity), the
tree walk necessary to build the interaction list is a relatively slow
operation. This is because it involves the evaluation of complex
conditional statements and requires ``pointer chasing'' operations. A
direct $N^2$ force calculation scales poorly as $N$ grows, but for a
small number of particles, a thoughtfully-constructed kernel can still
finish the computation in a small number of cycles. The RCB tree
exploits our highly-tuned short-range force kernels to decrease the
overall force evaluation time by shifting workload away from the slow
tree-walking and into the force kernel. Up to a point, doing this
actually speeds up the overall calculation: the time spent in the
force kernel goes up but the walk time decreases faster. Obviously, at
some point this breaks down, but on many systems, tens or hundreds of
particles can be in each leaf node before the crossover is reached. We
point out that the force kernel is generally more efficient as the
size of the interaction list grows: the relative loop overhead is
smaller, and more of the computation can be done using unrolled
vectorized code.

In addition to the performance benefits of grouping multiple particles
in each leaf node, doing so also increases the accuracy of the
resulting force calculation: The local force is dominated by nearby
particles, and as more particles are retained in each leaf node, more
of the force from those nearby particles is calculated exactly. In
highly-clustered regions (with very many nearby particles), the
accuracy can increase by several orders of magnitude when keeping over
100 particles per leaf node.

Another important consideration is the tree-node partitioning step,
which is the most expensive part of the tree build.  The particle data
is stored as a collection of arrays -- the so-called
structure-of-arrays (SOA) format. There are three arrays for the three
spatial coordinates, three for the velocity components, in addition to
arrays for mass, a particle identifier, etc.  Our implementation in
HACC divides the partitioning operation into three phases. The first
phase loops over the coordinate being used to divide the particles,
recording which particles will need to be swapped.  Next, these
prerecorded swapping operations are performed on six of the
arrays. The remaining arrays are identically handled in the third
phase. Dividing the work in this way allows the hardware prefetcher to
effectively hide the memory transfer latency during the particle
partitioning operation and reduces expensive read-after-write
dependencies.

\begin{figure}[t!]

  \centering \includegraphics[width=3.3in,angle=0]{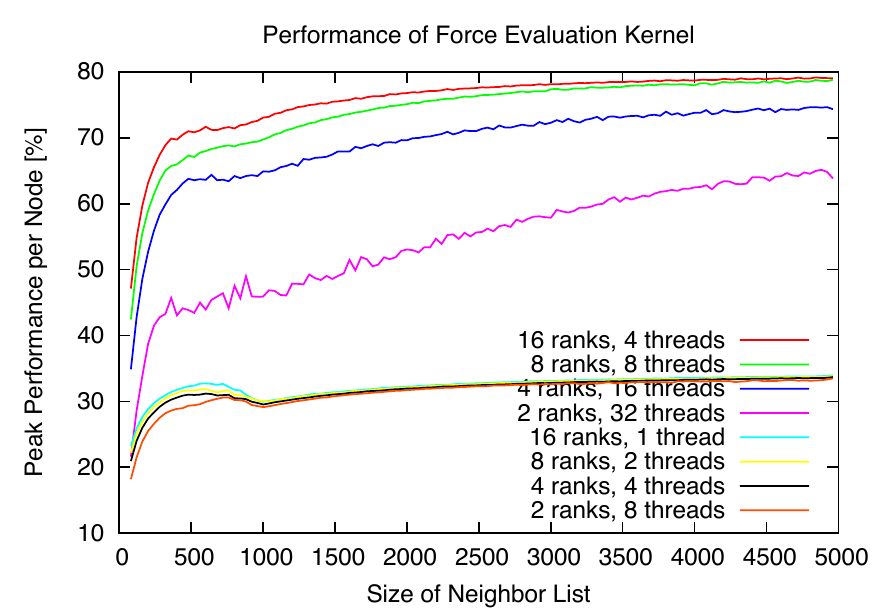}

  \caption{\small{\em Threading performance of the force evaluation
      kernel as a function of ranks per node and the total number of
      threads with varying neighbor list size (typical runs have
      neighbor list sizes $\sim$500-2500). At 4 threads/core, the
      performance attained is close to 80\% of peak. Note the
      exceptional performance even at 2 ranks per node.}}
\label{threads}
\end{figure}

We now turn to the evaluation of the BG/Q-specific short-range force
kernel, where the code spends the bulk of its computation time. Due to
the compactness of the short-range interaction
(Cf. Section~\ref{sec:hacc}), the kernel can be represented as
\begin{equation}
f_{SR}(s)=(s+\epsilon)^{-3/2}-f_{grid}(s)
\label{force}
\end{equation}
where $s={\mathbf r}\cdot{\mathbf r}$,
$f_{grid}(s)=\hbox{poly}[5](s)$, and $\epsilon$ is a short-distance
cutoff. This computation must be vectorized to attain high
performance; we do this by computing the force for every neighbor of
each particle at once. The list of neighbors is generated such that
each coordinate and the mass of each neighboring particle is
pre-generated into a contiguous array. This guarantees that 1) every
particle has an independent list of particles and can be processed
within a separate thread; and 2) every neighbor list can be accessed
with vector memory operations, because contiguity and alignment
restrictions are taken care of in advance. Every particle on a leaf
node shares the interaction list, therefore all particles have lists
of the same size, and the computational threads are automatically
balanced.

The filtering of $s$, i.e., checking the short-range condition, can be
processed during the generation of the neighbor list or during the
force evaluation itself; since the condition is likely violated only
in a number of ``corner'' cases, it is advantageous to include it into
the force evaluation in a form where ternary operators can be combined
to remove the need of storing a value during the force
computation. Each ternary operator can be implemented with the help of
the {\tt{fsel}} instruction, which also has a vector equivalent. Even
though these alterations introduce an (insignificant) increase in
instruction count, the entire force evaluation routine becomes fully
vectorizable.

On the BG/Q, the instruction latency is 6 cycles for most
floating-point instructions; latency is hidden from instruction
dependencies by a combination of placing the dependent instructions as
far as 6 instructions away by using 2-fold loop unrolling and
running 4 threads per core.

Register pressure for the 32 vector floating-point registers is the
most important design constraint on the kernel. Half of these
registers hold values common to all iterations, 6 of which store the
coefficients of the polynomial. The remaining registers hold
iteration-specific values.  Because of the 2-fold unrolling, this
means that we are restricted to 8 of these registers per
iteration. Evaluating the force in Eq.~(\ref{force}) requires a
reciprocal square root estimate and evaluating a fifth-order
polynomial, but these require only 5 and 2 iteration-specific
registers respectively. The most register-intensive phase of the
kernel loop is actually the calculation of $s$, requiring 3 registers
for the particle coordinates, 3 registers for the components of
${\mathbf r}$, and one register for accumulating the value of $s$.

There is significant flexibility in choosing the number of MPI ranks
versus the number of threads on an individual BG/Q node. Because of
the excellent performance of the memory sub-system and the low
overhead of context switching (due to use of the BQC wake-up unit), a
large number of OpenMP threads -- significantly larger than is
considered typical -- can be run to optimize
performance. Figure~\ref{threads} shows how increasing the number of
threads per core increases the performance of the force kernel as a
function of the size of the particle neighbor list. The best
performance is attained when running the maximum number of threads
(4) per core, and at large neighbor list size. 
The optimum
value for the current HACC runs turns out to be 16/4 as it allows for
short tree walks, efficient FFT computation, and a large fraction of
time devoted to the force kernel. The numbers in Fig.~\ref{threads}
show that for runs with different parameters, such as high particle
loading, the broad performance plateau allows us to use smaller or
higher rpn/thread ratios as appropriate.

The neighbor list sizes in representative simulations are of order
500-2500. At the lower end of this neighbor list size, performance can
be further improved using assembly-level programming, if desired.

At the chosen 16/4 operating point, the code spends 80\% of the time
in the highly optimized force kernel, 10\% in the tree walk, and 5\%
in the FFT, all other operations (tree build, CIC deposit) adding up
to another 5\%. Note that the actual fraction of peak performance
attained in the force kernel is close to 80\% as against a theoretical
maximum value of 81\%.  The 26 instructions in the kernel correspond
to a maximum of 208 Flops if they were all FMAs, whereas in the actual
implementation, 16 of them are FMAs yielding a total Flop count of
$168~(=40+128)$ implying a theoretical maximum value of $168/208=0.81$
or 81\% of peak.  The high efficiency is due to successful use of the
stream prefetch; we have measured the latency of the L2 cache to be
approximately 45 clock cycles, thus even a single miss per iteration
would be enough to significantly degrade the kernel performance.

\begin{figure}[b]
  \centering \includegraphics[width=2.8in,angle=0]{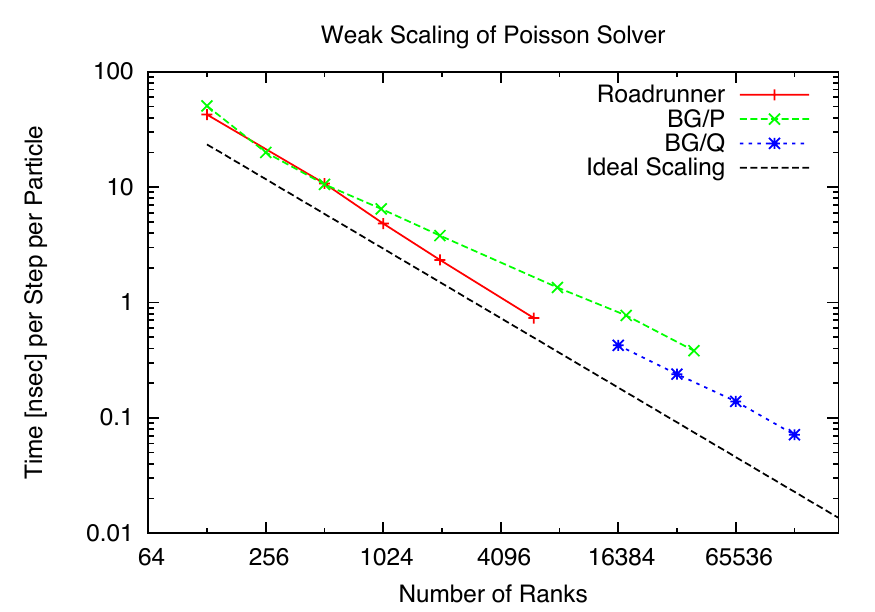}

\vspace{-0.2cm}

\caption{\small{\em Weak scaling of the Poisson Solver on different
    architectures. The Roadrunner scaling (red) is based on a
    slab-decomposed FFT, while the BG/P and BG/Q performance has been
    measured with the newer pencil-decomposed FFT. The black dashed
    line shows ideal scaling demonstrating the good performance of the
    long range solver.  }}
\label{poisson}
\end{figure}

\section{Performance}
\label{sec:results}

We present performance data for three cases: 1) weak scaling of the
Poisson solver up to 131,072 ranks on different architectures for
illustrative purposes; 2) weak scaling at $90\%$ parallel efficiency
for the full code on the BG/Q with up to 1,572,864 cores (96 racks);
and 3) strong scaling of the full code on the BG/Q up to 16,384 cores
with a fixed-size realistic problem to explore future systems with
lower memory per core. All the timing data were taken over many tens
of sub-steps, each individual run taking between 4-5 hours. 

To summarize our findings, both the long/medium range solver and the
full code exhibit perfect weak scaling out to the largest system we
have had access to so far; we achieved a performance of 13.94~PFlops
on 96 racks, at around 67-69\% of peak in most cases (up to
69.75\%). The full code demonstrates strong scaling up to one rack on
a problem with 1024$^3$ particles. Finally, the biggest test run
evolved more than 3.6~trillion particles (15,360$^3$), exceeding by
more than an order of magnitude, the largest high-resolution cosmology
run performed to date. As discussed in Section~\ref{science}, the
results of runs at this scale can be used for many scientific
investigations.

\subsection{Scaling of the Long/Medium-Range Solver}

\begin{table}
\renewcommand{\arraystretch}{1.5}
\caption{FFT Scaling on up to 10240$^3$ grid points on the BG/Q} 
\label{tab:fft}
\centering
\begin{tabular}{ccc}
FFT Size &  Ranks & Wall-clock Time [sec] \\ 
\hline
1024$^3$ & 256 & 2.731  \\
1024$^3$ & 512 &  1.392\\
1024$^3$ & 1024 &   0.713\\
1024$^3$ & 2048 &   0.354\\
1024$^3$ & 4096 &   0.179\\
1024$^3$ & 8192 &   0.098\\ 
\hline
4096$^3$ &16384 &  5.254\\
5120$^3$ & 32768 & 6.173 \\
6400$^3$ & 65536 & 6.841 \\
8192$^3$ & 131072 & 7.359 \\ 
9216$^3$ & 262144 & 7.238 \\
\hline
5120$^3$ &16384 &  10.36\\
6400$^3$ & 32768 & 12.40 \\
8192$^3$ & 65536 & 14.72 \\
10240$^3$ & 131072 & 14.24\\
\hline
\end{tabular}
\end{table}

As discussed earlier in Section~\ref{sec:hacc} the weak scaling
properties of HACC are controlled by the scaling properties of the
long/medium-range spectral force solver. The first version of HACC
used a slab-decomposed FFT, subject to the limit $N_{rank}<N_{FFT}$,
where $N_{rank}$ is the number of MPI ranks and the FFT is of size
$N_{FFT}^3$. In order to enable scalability to very large numbers of
cores, a fast and memory-efficient pencil-decomposed non-power-of-two
FFT (data partitioned across a 2-D subgrid) has been developed
allowing $N_{rank}<N_{FFT}^2$, sufficient for the foreseeable
future. The FFT is composed of interleaved transposition and
sequential 1-D FFT steps, where each transposition only involves a
subset of all tasks, and furthermore the transposition and 1-D FFT
steps can be overlapped and pipelined, with a reduction in
communication hotspots in the interconnect.  The details of the
implementation are rather complex, requiring a careful scheduling of
communication phases in order to avoid deadlock.

The scaling of the long/medium-range solver for both FFT
decompositions is shown in Fig.~\ref{poisson} for three different
architectures. On Roadrunner, the slab-decomposed FFT was used, while
on the BG/P and BG/Q, the pencil-decomposed FFT was used. In all
cases, the scaling is essentially ideal up to 131,072 ranks. The
largest FFT we ran for these tests had $N_{FFT}^3=10,240^3$ and a
run-time of less than 15~s.

\begin{table*}[t!]
\renewcommand{\arraystretch}{1.5}
\caption{Weak Scaling Performance on Mira and Sequoia
  ($\simeq$ 2,000,000 Particles per  Core), $N_p$=Number of particles,
  $L$=Boxlength.}   
\label{tab:perf1}
\begin{tabular}{ccccccccc}
Cores & $N_p$ & $L$ [Mpc] & Geom. & Total PFlops & Peak [\%]& Time/Substep/Particle [sec] & Cores*Time/Substep & Memory [MB/rank]\\ 
2,048 & 1600$^3$ & 1814 & 16x8x16       & 0.018 & 69.00 &  $4.12\cdot 10^{-8}$& $8.44\cdot 10^{-5}$ & 377 \\ 
4,096 & 2048$^3$ & 2286 & 16x16x16     & 0.036 & 68.59 & $1.92\cdot 10^{-8}$ & $7.86\cdot 10^{-5}$ & 380 \\
8,192 & 2560$^3$ & 2880 & 16x32x16     & 0.072 & 68.75 & $1.00\cdot 10^{-8}$ & $8.21\cdot 10^{-5}$ & 395 \\
16,384 & 3200$^3$ & 3628 & 32x32x16   & 0.144 & 68.50 & $5.19\cdot 10^{-9}$ & $8.50\cdot 10^{-5}$ & 376 \\
32,768 & 4096$^3$ & 4571 & 64x32x16   & 0.269 & 69.02 & $2.88\cdot 10^{-9}$ & $9.44\cdot 10^{-5}$ & 414 \\
65,536 & 5120$^3$ & 5714 & 64x64x16   & 0.576 & 68.64 & $1.46\cdot 10^{-9}$ & $9.59\cdot 10^{-5}$ & 418 \\
131,072 & 6656$^3$ & 6857 & 64x64x32 & 1.16 & 69.37 & $7.41\cdot 10^{-10}$ & $9.70\cdot 10^{-5}$ &377\\
262,144 & 8192$^3$ & 9142 & 64x64x64 & 2.27 & 67.70 & $3.04\cdot 10^{-10}$ & $7.96\cdot 10^{-5}$ & 346 \\
393,216 & 9216$^3$ & 9857 & 96x64x64 & 3.39 & 67.27 & $2.03\cdot 10^{-10}$ & $7.99\cdot 10^{-5}$ & 342 \\
524,288 & 10240$^3$ & 11429 & 128x64x64 & 4.53 & 67.46 & $1.59\cdot 10^{-10}$ & $8.36\cdot 10^{-5}$ & 348 \\
786,432 & 12288$^3$ & 13185 & 128x128x48 & 7.02 & 69.75 & $1.2\cdot 10^{-10}$  & $9.90\cdot 10^{-5}$&415\\
1,572,864 & 15360$^3$ & 16614 & 192x128x64 & 13.94 & 69.22 & $5.96\cdot 10^{-11}$ & $9.93\cdot 10^{-5}$&402
\end{tabular}
\end{table*}

More detailed timings of the pencil-decomposed FFT on the BG/Q system
are given in Table~\ref{tab:fft}. The first (top) part of the table
shows results from a strong scaling test for a fixed FFT size of
1024$^3$. As the number of ranks are increased from 256 to 8192 (one
rack of the BG/Q, 8 ranks per node), the scaling remains close to
ideal. In the second set of scaling tests, the grid size per rank is
held constant, at approximately 160$^3$. The FFT is scaled up to 24
racks and to a size of 9216$^3$. The performance is remarkably stable,
a successful benchmark for the BG/Q network. In the third set of
scaling tests, we increase the grid size per rank to approximately
200$^3$ per rank. The FFT scales up to 16 racks with a maximum size of
10240$^3$. These results predict excellent FFT performance on the
largest BG/Q systems available in the near future and beyond (as
demonstrated in the next sub-section).

\subsection{Scaling of the Full Code up to 96 Racks of the BG/Q} 

\begin{figure}[t!]
  \centering \includegraphics[width=3.5in,angle=0]{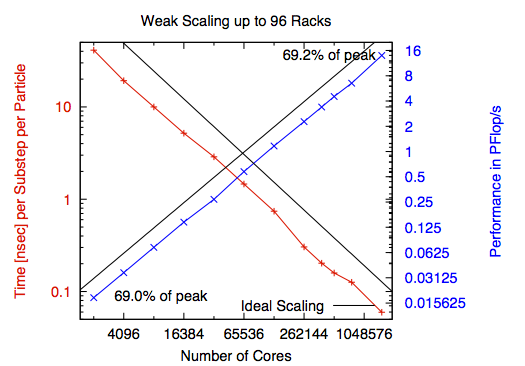}

\vspace{-0.2cm}

\caption{\small{\em Weak scaling for 2 million particles per core.
    The time per substep per particle (red) and the overall
    performance (blue) up to 96 racks of the BG/Q are shown as a
    function of the number of cores. The offset black dashed lines
    indicate ideal scaling. The performance and time to solution
    demonstrate essentially perfect scaling with the number of
    cores.}}
\label{weak}
\end{figure}

To demonstrate weak scaling of the full HACC framework we ran an
approximately fixed problem size with 2 million particles per core,
representative of the particle loading in actual large-scale
simulations. The results are shown in Fig.~\ref{weak} for both the
push-time per particle per substep (proportional to the wall-clock) as
well as for the total performance. Additional scaling tests for 4
million particles per core were carried out and showed very similar
performance. Table~\ref{tab:perf1} presents a more quantitative
picture of the results. The time to solution is set by the high-accuracy
science use requirement of running massive high-precision HACC
simulations on a production basis, i.e., within days rather than
weeks. Particle push-times of 0.06~ns/substep/particle for more than
3.6~trillion particles on 1,572,864 cores allow runs of 100 billion to
trillions of particles in a day to a week of wall-clock, and this
defines the approximate target. The results displayed in
Fig.~\ref{weak} show that we are achieving the required goal. The
largest problem was run with 3.6~trillion particles, more than a
factor of 10 larger than the Millennium XXL simulation~\cite{xxl},
currently the world's biggest high-resolution cosmological
simulation. (Lower resolution runs include examples at 374 billion
particles~\cite{kim11} and 550 billion particles~\cite{alimi12}, both
significantly smaller than our high-resolution benchmark.)

\begin{figure}[b!]
  \centering \includegraphics[width=3.3in,angle=0]{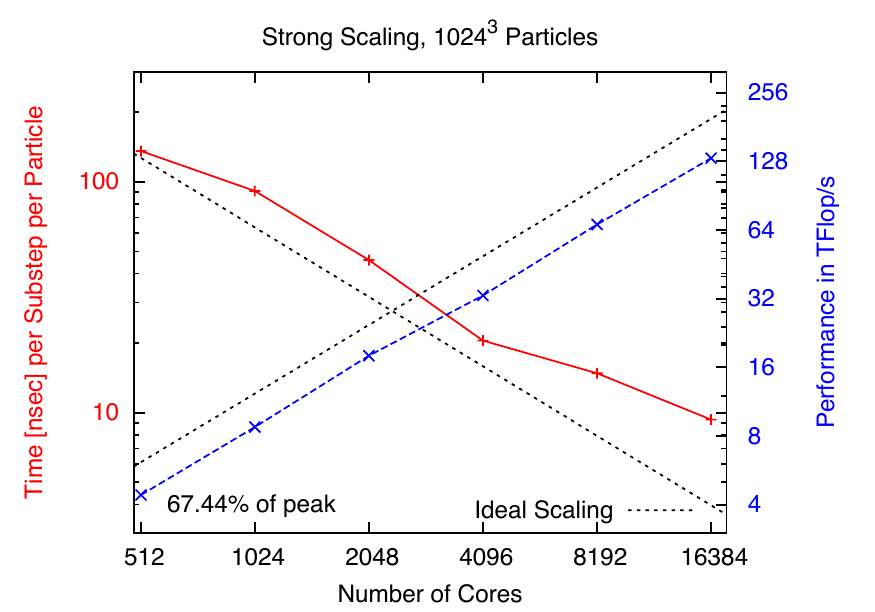}

\vspace{-0.2cm}

\caption{\small{\em Strong scaling for a fixed-size problem --
    1024$^3$ particles in a (1.42~Gpc)$^3$ box. The number of cores
    increases from 512 to 16384. Shown are the scaling of the time per
    step per particle (red) and the overall performance (blue).  The
    timing scales nearly perfectly up to 8192 cores, then slows down
    slightly. The performance stays high throughout. The typical
    particle loading per core for science runs will be between 2 and 4
    million particles (Cf. Fig.~\ref{weak}), corresponding to the 512
    core case in this figure. Successful scaling up to nearly the full
    rack (reducing particles/core to 65,536) is very encouraging in
    view of future platform constraints (see text for further
    discussion). These runs used an earlier version of the force
    kernel.}}
\label{strong}
\end{figure}

As demonstrated in Fig.~\ref{weak} and Table~\ref{tab:perf1}, weak
scaling is ideal up to 1,572,864 cores (96 racks), where HACC attains a
peak performance of 13.94~PFlops and a time per particle per substep of
$\sim 0.06$~ns for the full high-resolution code. We have single node
performance results for the entire 96 rack run, not just for the
kernel or time-stepping. The instruction mix is ${\rm FPU}=56.10\%$ and
${\rm FXU}=43.90\%$, therefore the maximal possible throughput is
$100/56.10=1.783$ instructions/cycle. The actual instructions/cycle
completed per core is 1.508, 85\% of the possible issue rate for our
instruction mix. The memory footprint per rank is 402~MB, or 6.4~GB
per node, therefore the data does not fit into L2 cache. The achieved
L1 hit rate is a remarkable 99.62\%, given the substantial memory
footprint. The counters report 142.32~GFlops from a 204.8~GFlops node,
i.e., 69.5\% of peak performance. The memory bandwidth is very low:
0.344~B/cycle out of a measured peak of 18~B/cycle; this testifies to
the very high rate of data reuse.

Our weak scaling performance results are encapsulated in
Table~\ref{tab:perf1} for up to 96 racks, the largest BG/Q system
worldwide. These results were obtained with code parameters set to
satisfy stringent accuracy requirements. Because of HACC's algorithmic
flexibility, time to solution can be balanced against accuracy as a
separate condition; we chose not to do that here.

\subsection{Strong Scaling of the Full Code on the BG/Q}

\begin{table*}[!t]
\renewcommand{\arraystretch}{1.5}
\caption{Strong Scaling Performance on Mira, 1024$^3$
  Particles total. Note: Results are from an earlier version of the kernel.} 
\label{tab:perf3}
\begin{tabular}{cccccccc}
 Cores & Particles/Core&Total TFlops & Peak [\%] & Time/Substep
 [sec] & Time/Substep/Particle [sec] & Memory [MB/rank]&Fraction of
 Memory [\%]\\ 
512    & 2,097,152 &  4.42 & 67.44 & 145.94 & $1.36\cdot 10^{-7}$ &
368.82 & 62.39 \\ 
1024  & 1,048,576 &  8.77 & 66.89 & 98.01  & $9.13\cdot 10^{-8}$ &
230.07 & 31.52\\ 
2048  & 524,288 & 17.99 & 68.67 &49.16 & $4.58\cdot 10^{-8}$ & 125.86
& 15.09\\ 
4096  & 262,144 & 33.06 & 63.05 & 21.97 & $2.05\cdot 10^{-8}$ &
75.816 &8.57\\ 
8192  & 131,072 & 67.72 & 64.59& 15.90 &  $1.48\cdot 10^{-8}$ & 57.15
& 6.33\\ 
16384& 65,536 & 131.27 & 62.59 & 10.01 & $9.33\cdot 10^{-9}$ & 41.355
& 4.50\\ 
\end{tabular}
\end{table*}

The evolution of many-core based architectures is strongly biased
towards a large number of (possibly heterogeneous) cores per compute
node. It is likely that the (memory) byte/flop ratio could easily
evolve to being worse by a factor of 10 than it is for the BG/Q, and
this continuing trend will be a defining characteristic of exascale
systems. For these future-looking reasons -- the arrival of the
strong-scaling barrier for large-scale codes -- and for optimizing
wall-clock for fixed problem size, it is important to study the
robustness of the strong scaling properties of the HACC
short/close-range algorithms.

We designed the test with a $1024^3$ particle problem running on
one rack from 512 to 16384 cores, spanning a per node memory
utilization factor of approximately 57\%, a typical production run
value, to as low as 7\%. The actual memory utilization factor scales
by a factor of 8, instead of 32, because on 16384 nodes we are running
a (severely) unrealistically small simulation volume per rank with
high particle overloading memory and compute cost. Despite this
`abuse' of the HACC algorithms, which are designed to run at $>50\%$
of per node memory utilization to about a factor of 4 less ($\sim
15\%$), the strong scaling performance, as depicted in Fig.~7
(associated data in Table~\ref{tab:perf3}) is impressive. The
performance stays near-ideal throughout, as does the push-time per
particle per step up to 8192 cores, slowing down at 16384 cores, only
because of the extra computations in the overloaded
regions. Therefore, we expect the basic algorithms to work extremely
well in situations where the byte/flop ratio is significantly smaller
than that of the current optimum plateau for the BG/Q.

\section{Early Scientific Results}
\label{science}

Dark energy is one of the most pressing puzzles in physics
today~\cite{frieman, DEreview}. HACC will be used to investigate the
signatures of different dark energy models in the detail needed to
analyze upcoming cosmological surveys. The cosmology community has
mostly focused on one cosmological model and a handful of `hero' runs
to study it~\cite{hubblevol, millennium, xxl}. With HACC, we aim to
systematically study dark energy model space at extreme scales and
derive not only qualitative signatures of different dark energy
scenarios but deliver quantitative predictions of unprecedented
accuracy urgently needed by the next-generation surveys. The
simulations can be used to interpret observations of various kinds,
such as weak gravitational lensing measurements to map the
distribution of dark matter in the Universe, measurements of the
distribution of galaxies and clusters, from the largest to the
smallest scales, measurements of the growth and distribution of cosmic
structure, gravitational lensing of the cosmic microwave background,
and many more.

\begin{figure*}{}

\centering \includegraphics[width=6.5in,angle=0]{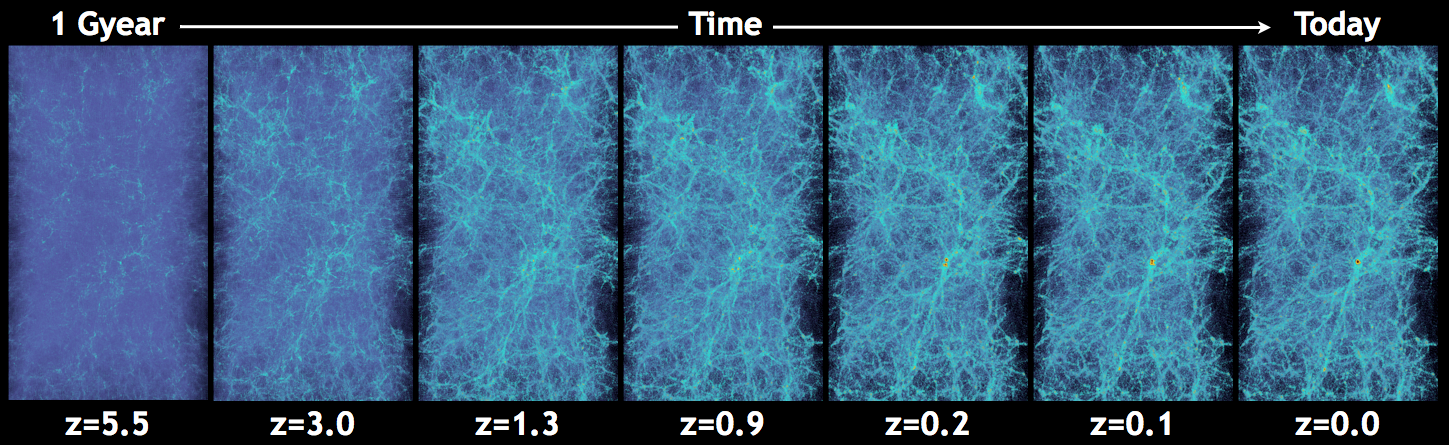}

\caption{\small{\em Time evolution of structure formation. A zoom-in
    to an approximately 70~Mpc wide region is shown. The frames depict
    the structure at different redshifts. Comparison to the overall
    box size of (9.14~Gpc)$^3$ shows the impressive dynamic range
    achievable on the BG/Q.\label{evolv}}}
\end{figure*}
We now show illustrative results from a science test run on 16 racks
of Mira, the BG/Q system now under acceptance testing at the ALCF. In
this simulation, 10240$^3$ particles are evolved in a (9.14~Gpc)$^3$
volume box. This leads to a particle mass, $m_p\simeq 1.9\cdot
10^{10}$~M$_\odot$, allowing us to resolve halos that host, e.g.,
luminous red galaxies (LRGs), a main target of the Sloan Digital Sky
Survey. The test simulation was started at an initial redshift of
$z_{in}=25$ (our actual science runs have $z_{in}\simeq 200$) and
evolved until today (redshift $z=0$). We stored a slice of the
three-dimensional density at the final time (only a small file system
was available during this run), as well as a subset of the particles
and the mass fluctuation power spectrum at 10 intermediate
snapshots. The total run time of the simulation was approximately 14
hours.  As the evolution proceeds, the particle distribution
transitions from essentially uniform to extremely clustered (see
Fig.~\ref{evolv}).  The local density contrast $\delta_m$ can increase
by five orders of magnitude during the evolution. Nevertheless, the
wall-clock per time step does not change much over the entire
simulation.

The large dynamic range of the simulation is demonstrated in
Fig.~\ref{zoom}. The outer image shows the full simulation volume
(9.14~Gpc on a side). In this case, structures are difficult to see
because the visualization cannot encompass the dynamic range of the
simulation. Successively zooming into smaller regions, down to a
(7~Mpc)$^3$ sub-volume holding a large dark matter-dominated halo
gives some impression of the enormous amount of information contained
in the simulation. The zoomed-in halo corresponds to a cluster of
galaxies in the observed Universe. Note that in this region the actual
(formal) force resolution of the simulation is 0.007~Mpc, a further
factor of 1000 smaller than the sub-volume size!

A full simulation of the type described is extremely science-rich and
can be used in a variety of ways to study cosmology as well as to
analyze available observations. Below we give two examples of the kind
of information that can be gained from large scale structure
simulations. (We note that the test run is more than three times
bigger than the largest high-resolution simulation available today.)

Clusters are very useful probes of cosmology -- as the largest
gravitationally bound structures in the Universe, they form very late
and are hence sensitive probes of the late-time acceleration of the
Universe~\cite{Vikhlinin,Haiman}. Figure~\ref{cluster} gives an
example of the detailed information available in a simulation,
allowing the statistics of halo mergers and halo build-up through
sub-halo accretion to be studied with excellent statistics. In
addition, the number of clusters as a function of their mass (the mass
function), is a powerful cosmological probe. Simulations provide
precision predictions that can be compared to observations. The new
HACC simulations will not only predict the mass function (as a
function of cosmological models) at unprecedented accuracy, but also
the probability of finding very massive clusters in the
Universe. Large-volume simulations are required to determine the
abundance of these very rare objects reliably~\cite{Harrison,
  Mortonson, elgordo}.

\begin{figure}[h!]
  \centering \includegraphics[width=3.3in,angle=0]{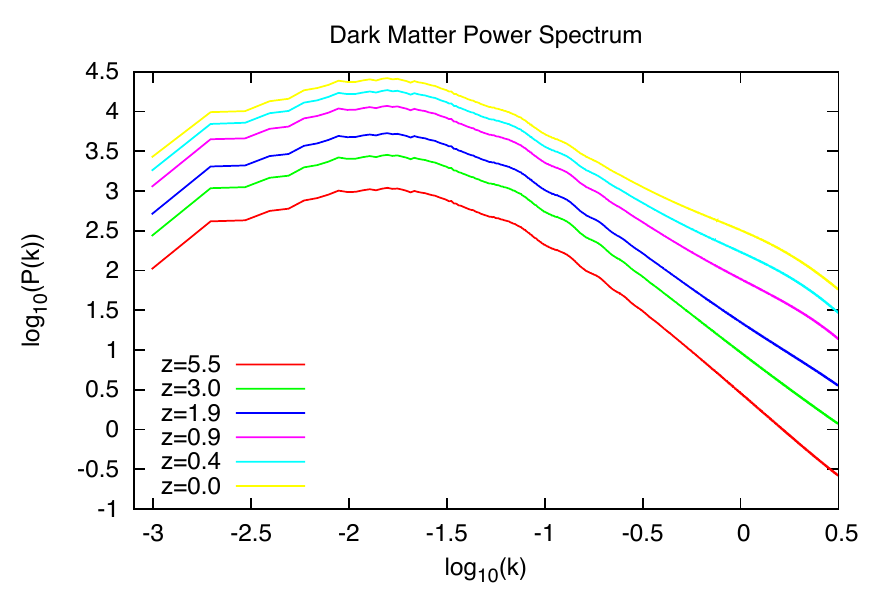}

\vspace{-0.2cm}

\caption{\small{\em Evolution of the matter fluctuation power spectrum.}}
\label{power}
\end{figure}

\begin{figure}[t!]
  \centering \includegraphics[width=2.8in,angle=0]{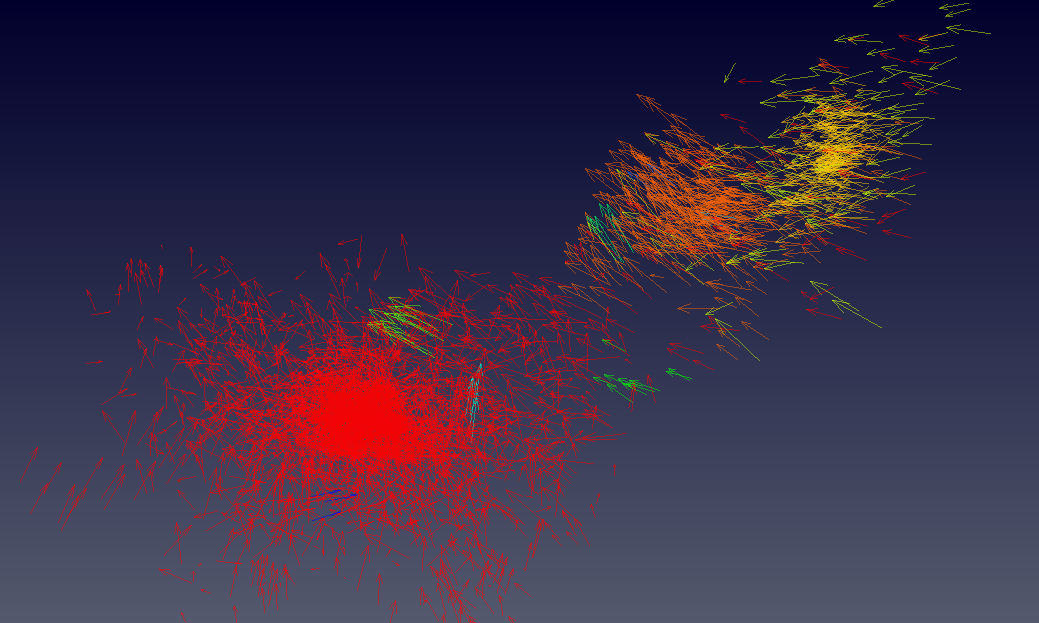}
  \centering \includegraphics[width=2.8in,
  height=2in,angle=0]{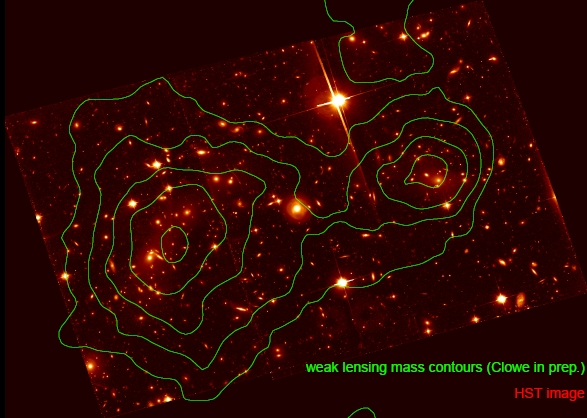}

\vspace{-0.2cm}

\caption{\small{\em Upper panel: Visualization of a large halo of mass
    $\simeq10^{15}~${\rm M}$_\odot$ from a simulation (similar to
    the zoom-in region in Fig.~\ref{zoom}) including sub-halos. Each
    sub-halo is shown in a different color. Only a fraction of the
    tracer particles are shown for better visibility -- the particles
    are shown as arrows, directed along their velocity vectors. The
    main halo (red) is in a relatively relaxed configuration; it will
    host a bright central galaxy as well as tens of dimmer
    galaxies. Each sub-halo, depending on its mass, can host one or
    more galaxies. Lower panel: Optical observations and weak lensing
    contours (green) of a cluster~\cite{clowe}. The contours trace the
    matter density and can be directly compared to simulation
    outputs. As in the simulation example, this cluster has two major
    mass components.}}
\label{cluster}
\end{figure}

Cosmological information resides in the nature of material structure
and also in how structures grow with time. To test whether general
relativity correctly describes the dynamics of the Universe,
information related to structure growth (evolution of clustering) is
essential. Figure~\ref{evolv} shows how structure evolves in the
simulation. Large-volume simulations are essential in producing
predictions for statistical quantities such as galaxy correlation
functions and the associated power spectra -- with small statistical
errors -- in order to compare the predictions against
observations. Figure~\ref{power} shows how the power spectrum evolves
as a function of time in the science test run. At small wavenumbers,
the evolution is linear, but at large wavenumbers it is highly
nonlinear, and cannot be obtained by any method other than direct
simulation.

To summarize, armed with large-scale simulations we can study and
evaluate many cosmological probes. These probes involve the
statistical measurements of the matter distribution at a given epoch
(such as the power spectrum and the mass function) as well as their
evolution. In addition, the occurrence of rare objects such as very
massive clusters can be investigated in the simulations we will carry
out with HACC.

\section{The Future}
\label{sec:future}

These are exciting times for users of the BG/Q: In the US, two large
systems are undergoing acceptance at Livermore (Sequoia, 96 racks) and
at Argonne (Mira, 48 racks). As shown here, HACC scales easily to 96
racks. Our next step is to exploit the power of the new systems with
the aim of carrying out a suite of full science runs with hundreds of
billions to trillions of simulation particles.

Because HACC's performance and scalability do not rely on the use of
vendor-supplied or other `black box' high-performance libraries or
linear algebra packages, it retains the key advantage of allowing code
optimization to be a continuous process; we have identified several
options to enhance the performance as reported here. An initial step
is to fully thread all the components of the long-range solver, in
particular the forward CIC algorithm. Next, we will improve (nodal)
load balancing by using multiple trees at each rank, enabling an
improved threading of the tree-build. While the force kernel already
runs at very high performance, we can improve it further with
lower-level implementations in assembly.

Many of the ideas and methods presented here are relatively general
and can be re-purposed to benefit other HPC applications. In addition,
HACC's extreme continuous performance contrasts with the more bursty
stressing of the BG/Q architecture by Linpack; this feature has allowed
HACC to serve as a valuable stress test in the Mira and Sequoia
bring-up process.

To summarize, we have demonstrated outstanding performance at close to
14~PFlops on the BG/Q (69\% of peak) using more than 1.5 million cores
and MPI ranks, at a concurrency level of 6.3 million. We are now ready
to carry out detailed large-volume N-body cosmological simulations at
the size scale of trillions of particles.

\section*{Acknowledgment}
We are indebted to Bob Walkup for running HACC on a prototype BG/Q
system at IBM and to Dewey Dasher for help in arranging access. At
ANL, we thank Susan Coghlan, Paul Messina, Mike Papka, Rick Stevens,
and Tim Williams for obtaining allocations on different Blue Gene
systems. At LLNL, we are grateful to Brian Carnes, Kim Cupps, David
Fox, and Michel McCoy for providing access to Sequoia. We thank Ray
Loy and Venkat Vishwanath for their contributions to system
troubleshooting and parallel I/O. We acknowledge the efforts of the
ALCF operations team for their assistance in running on Mira and the
VEAS BG/Q system, in particular, to Paul Rich, Adam Scovel, Tisha
Stacey, and William Scullin for their tireless efforts to keep Mira
and VEAS up and running and helping us carry out the initial
long-duration science test runs. This research used resources of the
ALCF, which is supported by DOE/SC under contract DE-AC02-06CH11357. 

\clearpage

\bibliographystyle{plain}

\bigskip

\framebox{
  \parbox{3in}{ The submitted manuscript has been created by UChicago
    Argonne, LLC, Operator of Argonne National Laboratory
    (``Argonne"). Argonne, a U.S. Department of Energy Office of
    Science laboratory, is operated under Contract
    No. DE-AC02-06CH11357. The U.S. Government retains for itself, and
    others acting on its behalf, a paid-up nonexclusive, irrevocable
    worldwide license in said article to reproduce, prepare derivative
    works, distribute copies to the public, and perform publicly and
    display publicly, by or on behalf of the Government.  }
}

\end{document}